\documentclass[reprint,pre,superscriptaddress,floatfix,showpacs]{revtex4-1}
\usepackage{graphicx}
\usepackage{rotating}
\usepackage{amsfonts,amsmath,color}
\usepackage{amssymb}
\usepackage{psfrag}

\newcommand{\bs}{{\bf {s}}}
\newcommand{\br}{{\bf {r}}}
\newcommand{\bv}{{\bf {v}}}

\begin{document}

\title{Three-dimensional inverse energy transfer induced by vortex
reconnections}

\author{Andrew W. Baggaley}
\email{andrew.baggaley@glasgow.ac.uk}
\affiliation{
School of Mathematics and Statistics, University of Glasgow,
Glasgow, G12 8QW, UK
}
\author{Carlo F. Barenghi}
\email{c.f.barenghi@ncl.ac.uk}
\affiliation{
Joint Quantum Centre Durham-Newcastle, and 
School of Mathematics and Statistics, Newcastle University,
Newcastle upon Tyne, NE1 7RU, UK
}

\author{Yuri A. Sergeev}
\email{yuri.sergeev@ncl.ac.uk}
\affiliation{
Joint Quantum Centre Durham-Newcastle, and 
School of Mechanical and Systems Engineering,
Newcastle University, 
Newcastle upon Tyne, NE1 7RU, UK
} 


\begin{abstract}
In low-temperature superfluid helium,
viscosity is zero, and vorticity takes the form of discrete, 
vortex filaments of fixed circulation and atomic thickness.
We present numerical evidence of 
three-dimensional inverse energy transfer from small length scales to large
length scales in superfluid turbulence generated by a
flow of vortex rings. We argue that the effect
arises from the anisotropy of the flow, which favours
vortex reconnections of vortex loops of the same polarity,
and that it has been indirectly observed in the laboratory.
The effect open questions about analogies with related processes
in ordinary turbulence.

\end{abstract}

\maketitle

The phenomenology of three-dimensional
turbulence is based on Richardson's idea \cite{Richardson} of the 
(forward) turbulent cascade. Kinetic energy, injected
externally at large length scales, feeds large unstable eddies, 
which interact, become stretched and break up
into smaller eddies. The process repeats, until, at
sufficiently small length scales, viscous forces dissipate 
energy into heat.
A reversed flux of energy, from the small scales to
the large scales, is observed in two-dimensional turbulence
\cite{Kraichnan-Montgomery,Tabeling-2D}.
Such inverse cascade is more rare in three-dimensional turbulence,
but can be observed in the presence of strong anisotropy
\cite{Galanti-Sulem,Hefer-Yakhot,Yakhot-Pelz,Yakhot-Sivashinsky},
or when rotation \cite{Mininni-Pouquet,Pouquet-2012,Smith-Waleffe}
or stratification \cite{Xia-Falko} make the flow almost
two-dimensional. 
An example is Jupiter's Great Red Spot \cite{Marcus,Swinney}.

At temperatures below $1~\rm K$, thermal excitations can be neglected
and liquid helium ($^4$He) is a pure superfluid. 
Unlike ordinary fluids (in which vorticity is a continuous field),
the superfluid's rotational motion is constrained by quantum mechanics to 
discrete vortex lines of fixed circulation 
$\kappa$ and atomic thickness (the
radius of the vortex core is only $a_0 \approx 10^{-8}~\rm cm$). 
Turbulence \cite{Skrbek-Sreeni}, easily excited by stirring the
liquid helium, is a tangle of such vortex lines. An important property
of vortex lines is that they reconnect when they come 
sufficiently close to each other, as predicted by theory 
\cite{Koplik,Bajer,Zuccher} and observed in experiments \cite{Paoletti}.
Superfluid reconnections are similar to reconnections in ordinary 
fluids \cite{Kerr-dublin}.

In this report, we exploit the singular nature of 
superfluid vorticity to examine the three-dimensional inverse 
energy transfer. 
Using numerical simulations, firstly we demonstrate  
that an inverse energy transfer is possible
in superfluid helium (and, we argue, it has already been
observed in the laboratory, although indirectly). Secondly, we show that vortex
reconnections play a key role in this process. 

We numerically model vortex lines \cite{Schwarz}
as oriented space curves $\bs(\xi,t)$ of infinitesimal 
thickness, where $\xi$ is arc length and $t$ is time. This approach
is justified by the large separation of scales between $a_0$ and
the typical distance between vortices, $\ell \approx L^{-1/2}$
(where $L=\Lambda/V$ is the vortex line density, $\Lambda$ the
vortex length and $V$ volume).
Two physical ingredients determine the evolution of vortex lines. 
The first  is Helmoltz's theorem: a vortex at location $\bs$
is swept by velocity field $\bv$ generated by the entire vortex
configuration $\cal L$ at $\bs$ via
the Biot-Savart law \cite{Saffman}:

\begin{equation}
\frac{d \bs}{dt}=\bv(\bs,t),
\qquad
\bv(\bs,t)=
-\frac{\kappa}{4 \pi} \oint_{\cal L} \frac{(\bs-\br) }
{\vert \bs - \br \vert^3}
\times {\bf d}\br,
\label{e:BS}
\end{equation}
\noindent
where $\kappa=9.97 \times 10^{-4}~\rm cm^2/s$.
The second ingredient, mentioned before, is vortex reconnections,
instantaneous events which occur when vortex lines
collide. 

Our numerical simulations are performed in a periodic cube
of size $D$.
The techniques to discretize vortex lines
into a variable number of points held
at minimum separation $\delta/2$,
time-step Eq.~(\ref{e:BS}), de-singularize  the
Biot-Savart integrals Eq.~(\ref{e:BS}) (right), and evaluate them 
via a tree-method \cite{angle}
are described in the literature \cite{Baggaley-fluctuations,Baggaley-tree}.
The reconnection algorithm is described in 
\cite{Baggaley-reconnections} and compared to other 
published algorithms.

In a pure superfluid, although viscosity is zero, 
the kinetic energy $K(t)$ is not conserved, but is turned into sound
(phonons) by rapidly rotating Kelvin waves (helical perturbations 
of vortex lines) at length scales of the order of $10^2~a_0$ \cite{Vinen2001}. 
In our simulations it is impossible to discretize 
vortex lines down to almost the atomic scale. However, the finite
numerical resolution qualitatively models phonon losses
\cite{Baggaley-cascade}, because
it damps out Kelvin waves at scales of the order 
of $\delta$, and slightly reduces the
vortex length (again at scale $\delta$) at each reconnection event.
Kelvin waves are often studied in the context of the
decay of superfluid turbulence at very low temperatures. 
Interacting Kelvin waves (see \cite{Krstulovic} and references
therein) form a one-dimensional weakly nonlinear system
in which a dual cascade in k-space takes place:
a direct cascade of energy to large $k$ and an inverse cascade of
wave action to small $k$ \cite{Nazarenko}.
Generation of long waves as well as short waves on individual vortices has been
observed in numerical simulations \cite{Vinen-cascade}, but this Kelvin
cascade process is not directly relevant to the inverse energy
transfer which we present here, which is three-dimensional in nature.

\begin{figure}
 \begin{center}
 \includegraphics[width=0.32\textwidth]{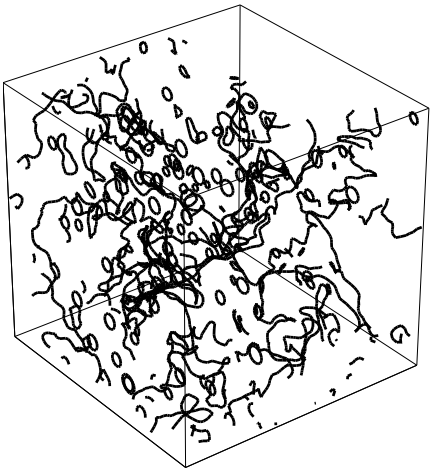}
 \end{center}
 \caption{
 First numerical simulation.  Snapshots of the vortex tangle in the
 steady--state regime at $t=4~\rm s$ ($L \approx 6000~\rm cm^{-2}$).
 }
 \label{fig:1}
 \end{figure}

\begin{figure}
\begin{center}
\includegraphics[width=0.5\textwidth]{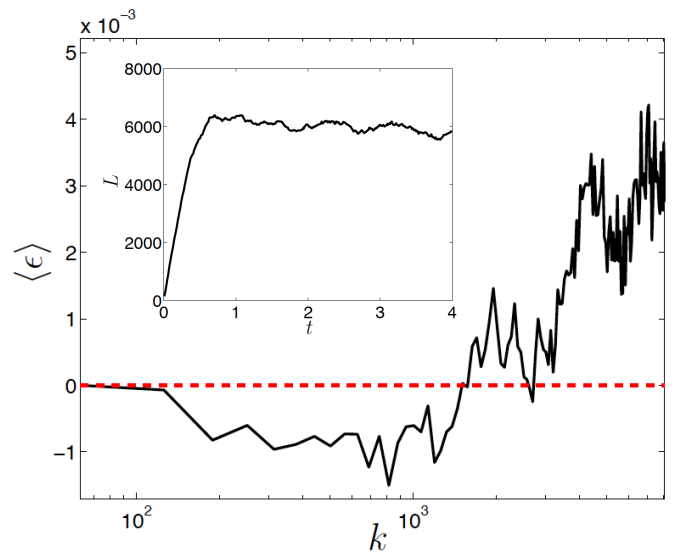}
\end{center}
\caption{(Color online). First numerical simulation.
Energy flux $\langle \epsilon \rangle$ 
(averaged over the statistical steady regime $2 <t<4~\rm s$)
vs wavenumber $k$ ($\rm cm^{-1}$). Notice that
$\langle \epsilon \rangle <0$ for $k<k_f$ (energy is transferred from
small to large length scales), and that $\langle \epsilon \rangle >0$
for $k>k_f$ (energy is transferred from large to small scales),
where $k_f \approx 1300~\rm cm^{-1}$ is the wavenumber corresponding
to the average diameter of the injected rings.
The inset shows the vortex line density $L$ ($\rm cm^{-2}$) vs time $t$
($\rm s$). 
}
\label{fig:2}
\end{figure}

Our first numerical
simulation \cite{sim1} models experiments \cite{Walmsley-Golov}. 
We start with an empty
computational box and inject vortex rings at frequency $f$
drawing their radius from a normal distribution.  
The rings are in the $yz$-plane and travel in the 
positive $x$ direction; their evolution is computed using the full
Biot-Savart law and the reconnection algorithm.
After an initial transient, the vortex system settles to a  
statistical steady state.
A snapshot of the vortex tangle is shown
in Fig.~\ref{fig:1}. In this regime
forcing is balanced by dissipation, and the vortex line
density fluctuates about a saturated value, as shown in
the inset of Fig.~\ref{fig:2}.

To determine the distribution of the kinetic energy over the length scales,
we Fourier--transform \cite{spectrum}
the velocity field $\bv$ and define the energy spectrum $E(k)$ by

\begin{equation}
K(t)=\frac{1}{V} \int \frac{1}{2}{\bv}^2 dV=\int_0^{\infty} E(k) dk,
\label{e:spectrum}
\end{equation}

\noindent
where $k=\vert \bf k \vert$ is the magnitude of the three-dimensional
wavenumber and $V=D^3$ is volume. Large (small) length scales correspond
to small (large) wavenumbers $k$ respectively. We find that, during
the evolution, $E(k)$ progressively increases at small $k$.
To quantify this energy transfer to large length scales,
we compute the energy flux $\epsilon(k)=-\int_{k_D}^k d E(k')/dt~dk'$
with $k_D=2 \pi/D$. Fig.~\ref{fig:2} shows that, in the 
statistically steady regime, the time-averaged energy
flux $\langle \epsilon \rangle$ is negative for $k<k_f$ and 
positive for $k>k_f$, where
$k_f=2 \pi / (2 \bar{R}) \approx 1300~\rm cm^{-1}$ is the forcing wavenumber
based on the injected rings' mean diameter $2 \bar{R}$.

\begin{figure}
\begin{center}
\includegraphics[width=0.5\textwidth]{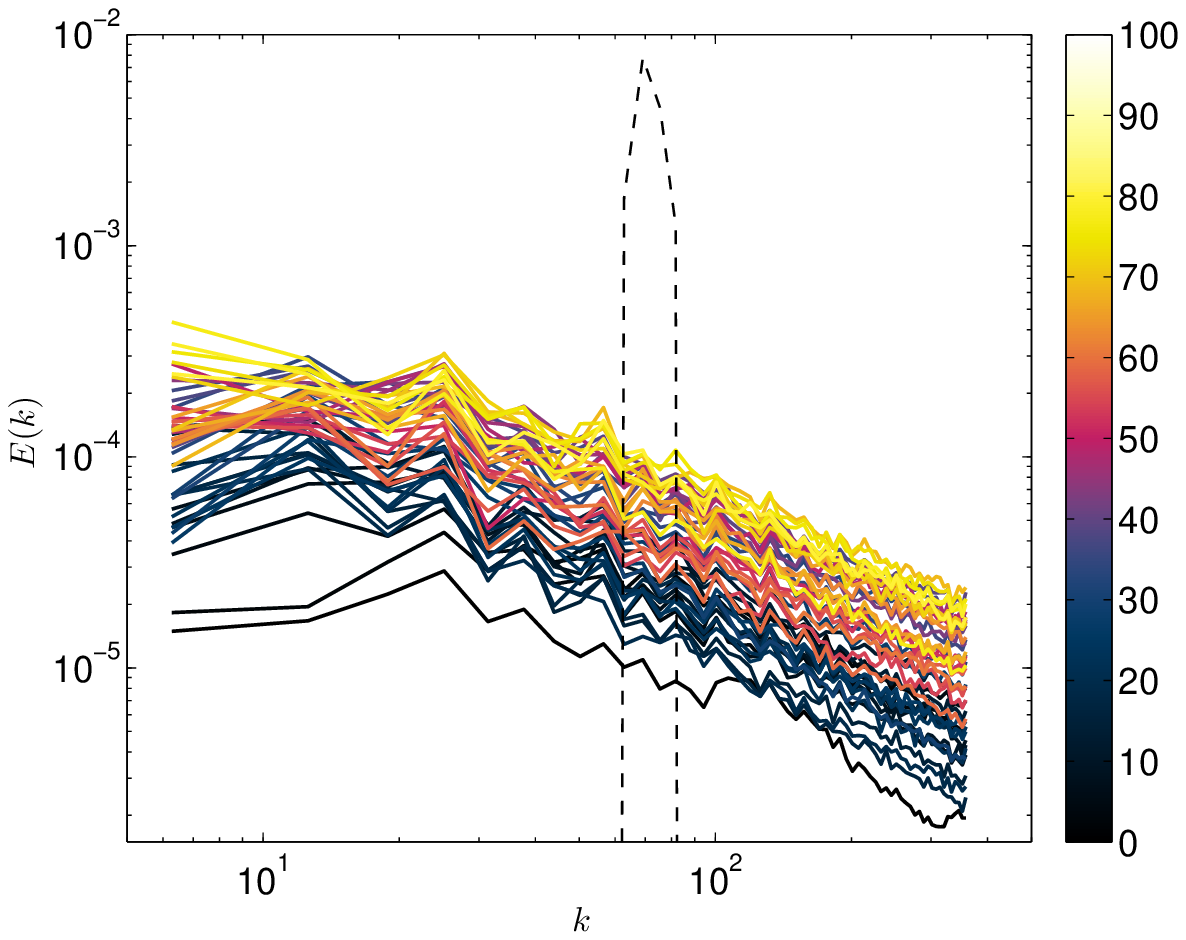}
\end{center}
\caption{(Color online). Second numerical simulation.
Energy spectra $E(k)$ (arbitrary units)
vs wavenumber $k$ ($\rm cm^{-1}$) shaded according to time $t$
($\rm s$) as in the legend.
The dashed line is the spectrum (arbitrary units, scaled for visibility)
of the forcing term $\bv_{ext}$ with the maximum at $k=k_f$. Note the growth of the spectrum
at all wavenumbers, particularly for $k<k_f$. 
}
\label{fig:3}
\end{figure}

In other numerical simulations we examine the inverse energy transfer
under different conditions.
The second simulation \cite{sim2} is inspired by  numerical studies of
homogeneous isotropic turbulence in which
a forcing term is added to the governing
Navier--Stokes equation to balance viscous dissipation and achieve
a statistically-steady state, independent of the
initial condition.
We add a random, incompressible, isotropic velocity field $\bv_{ext}$
to the right-hand-side of Eq.~\ref{e:BS}(left), consisting of
100 random Fourier modes, narrow
banded ($\Delta k \approx 2~\rm cm^{-1}$) around wavenumber
$k_f\approx 70~\rm cm^{-1}$, with
$\langle \bv^2_{ext} \rangle^{1/2}=3.1~\rm cm/s$.
The seeding initial condition consists of a small number
of randomly oriented vortex rings.
During the evolution,
energy is fed into the system by $\bv_{ext}$ and removed by the
numerical dissipation.
Fig.~\ref{fig:3} shows the growth
of $E(k)$ during the initial evolution
and the overall build up of $K(t)$ (the area under $E(k)$).
The dashed line is the spectrum of the forcing term $\bv_{ext}$.
The transfer of energy from large $k$ near the forcing $k_f$
to small $k$ is apparent: at small $k$,
$E(k)$ grows by a factor of 500.

\begin{figure}
\begin{center}
\includegraphics[width=0.5\textwidth]{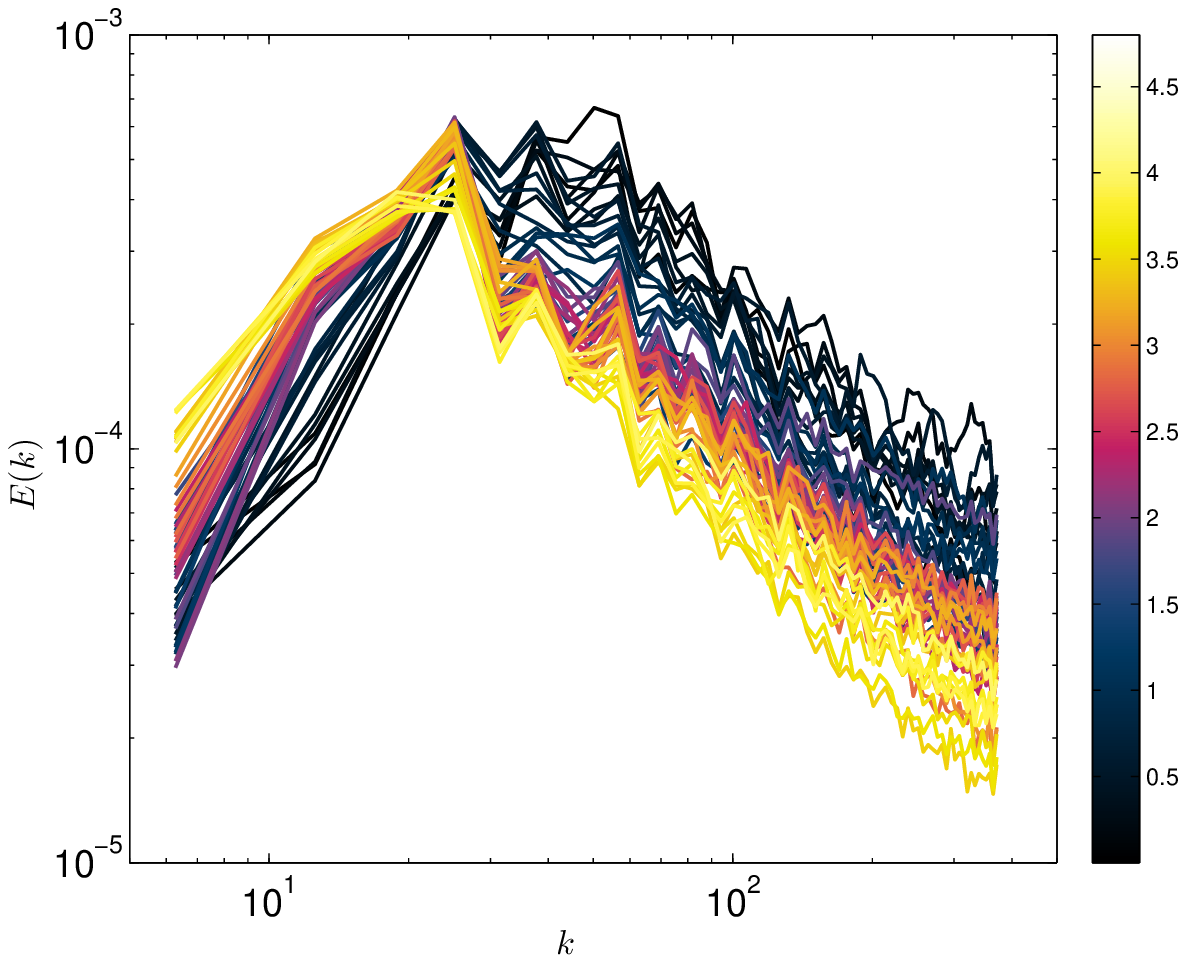}\\
\end{center}
\caption{(Color online).
Third simulation. Evolution of energy spectrum $E(k)$ vs wavenumber $k$.
The set-up is as in the first simulation (vortex rings of random
radius and travelling in the $x$ direction are injected),
but LIA replaces the Biot-Savart law. In this way
vortex lines interact only when they reconnect. 
Note again the transfer of energy from large $k$ to small $k$.
}
\label{fig:4}
\end{figure}

In the third numerical
simulation \cite{sim3} we examine the role of reconnections.
We proceed as in the first simulation, injecting rings of random
radius aligned in the $yz$ plane and travelling in the $x$ direction;
we retain the reconnection algorithm,
but replace the Biot-Savart law (Eq.~\ref{e:BS} (right)) with its 
Local Induction Approximation (LIA) \cite{DaRios,Ricca-darios}:

\begin{equation}
\bv(\bs,t) \approx \frac{\kappa}{4 \pi} \ln{(R/a_0)} \bs ' \times \bs'',
\label{eq:LIA}
\end{equation}

\noindent
where a prime denotes derivative with respect to arc length and
$R=1/\vert \bs'' \vert$ is the local radius of curvature.
Under LIA, vortex lines move along the binormal direction
with speed inversely proportional to $R$, ignoring
each other; in other words, vortices interact only when they collide.
Fig.~\ref{fig:4} shows that, in the absence of forcing,
$K(t)$ decreases, but energy is shifted from large $k$ to small $k$,
as in the previous simulations. This result means
that vortex-vortex interaction (represented by the Biot-Savart law) is not
necessary to produce a reverse energy transfer: 
vortex reconnections are enough to 
drive the process.

\begin{figure}
\begin{center}
\includegraphics[width=0.50\textwidth]{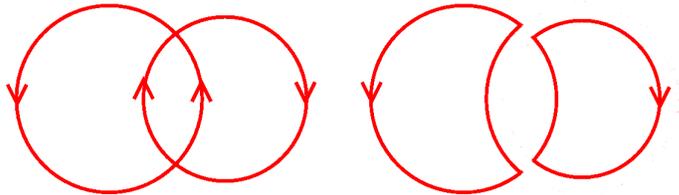}\\
\end{center}
\caption{(Color online).
Schematic head-on collision of vortex loops travelling in opposite
direction. After reconnection, the loops have essentially the same size.
}
\label{fig:5}
\end{figure}

\begin{figure}
\begin{center}
\includegraphics[width=0.50\textwidth]{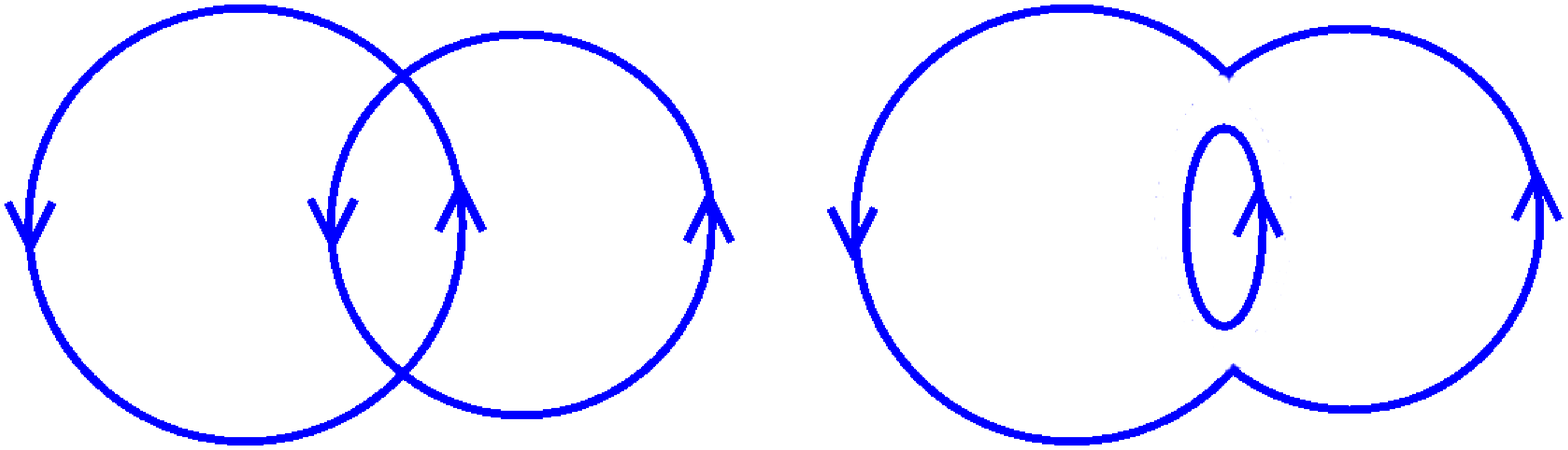}
\end{center}
\caption{(Color online).
Schematic collision  of loops
travelling in the same direction. After the reconnection, the loops have
very different size.
}
\label{fig:6}
\end{figure}

A simple geometrical interpretation of this result is the following.
Energy and speed of a vortex loop of size $R$ are roughly proportional to 
$R$ and $1/R$ respectively. Vortex loops travelling parallel or antiparallel
to a given direction undergo two kinds of collisions: head-on and from behind.
Head-one collisions leave the size of loops approximately unchanged 
after the reconnection, as in Fig.~\ref{fig:5}; collisions
from behind create a larger
loop and a smaller loop, as in Fig.~\ref{fig:6}. The large loop, which
contains most of the energy, is more likely to become entangled with
other vortices, while the small loop, which quickly moves
away, is more likely to be absorbed by walls 
(in the presence of periodic boundary conditions, the small loops
which collide with larger loops can only become smaller, without
significantly increasing the size of the larger loops).
In an isotropic tangle, collisions
of either kind are equally likely. In an anisotropic system (a jet of rings,
or a system in which loops are injected along a preferred direction),
we expect more collisions from behind, particularly in the early stage, hence
a shift of energy to larger length scales induced by reconnections alone. 

\begin{figure}
 \begin{center}
 \includegraphics[width=0.32\textwidth]{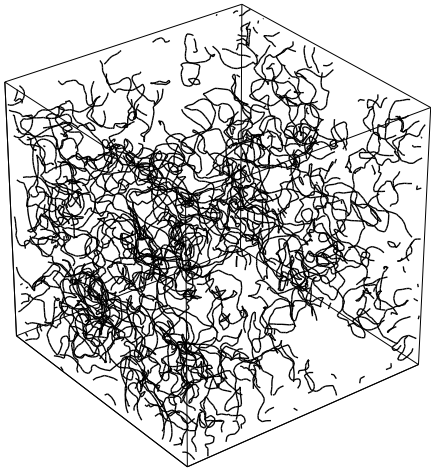}
 \end{center}
 \caption{
Fourth numerical simulation.
Snapshots of the vortex tangle in the steady state regime at
$t=15~\rm s$
($L \approx 160~\rm cm^{-2})$.
}
\label{fig:7}
 \end{figure}

\begin{figure}
 \begin{center}
 \includegraphics[width=0.5\textwidth]{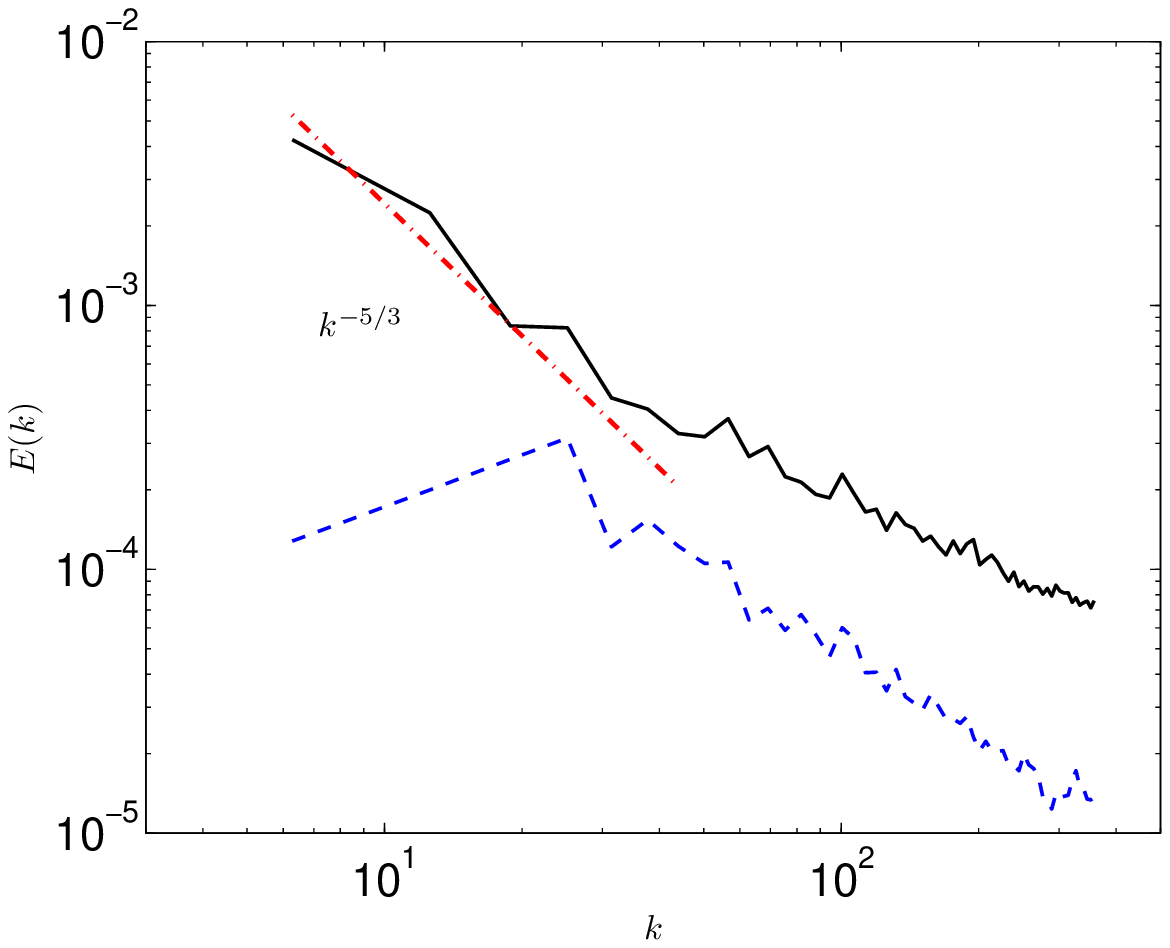}
 \end{center}
 \caption{(Color online)
Fourth simulation. Energy spectrum $E(k)$ (arbitrary units) vs 
wavenumber $k$ ($\rm cm^{-1}$)
at $t=0.75~\rm s$, $L=43.9~\rm cm^{-2}$, (dashed blue line) 
and $t=15~\rm s$, $L=159.6~\rm cm^{-2}$,  (solid black line). 
The dot-dashed line shows the $k^{-5/3}$ Kolmogorov scaling.
}
 \label{fig:8}
 \end{figure}

In the fourth simulation \cite{sim4} we proceed as in the first:
we inject rings moving in the $x$ direction
continually (drawing their radius from a normal distribution 
and compute the evolution using the Biot-Savart law and the 
reconnection algorithm); the difference is that now the forcing is 
relatively larger than in the first simulation~\cite{forcing}.
When the vortex line density saturates, the tangle settles down to a 
steady state 
(a snapshot of the saturated vortex tangle is shown in
Fig.~\ref{fig:7}).
We notice (see Fig.~\ref{fig:8})
that so much energy has been shifted to wavenumbers smaller
than the injection's inverse lengthscale $k_f \approx 1/\bar{R}$ 
that the spectrum
has acquired a form which is consistent with the classical Kolmogorov
scaling $E(k) \sim k^{-5/3}$ typical of ordinary turbulence, a result
which is in agreement with existing numerical simulations of superfluid
turbulence~\cite{Nore,Araki,Kobayashi,Sasa,Baggaley-structures}.


Now we put these numerical results in the context of experiments.
Walmsley \& Golov \cite{Walmsley-Golov} created turbulence in $^4$He
at very low temperatures (so that the normal fluid can be neglected)
by injecting vortex rings with a high voltage tip.
After the injection stage, they monitored the decay of the vortex line 
density $L$ and observed two 
regimes, $L \sim t^{-1}$ (called ``ultraquantum'') and $L \sim t^{-3/2}$
(called ``quasiclassical''), 
associated with short and long initial injection times
respectively. Both regimes were also observed
in $^3$He-B \cite{Bradley2006}.
In our previous paper \cite{Baggaley-ultra} we modelled 
the experiment of Walmsley \& Golov as realistically as possible, 
numerically injecting vortex rings
in the form of a narrow beam originating from a point source.
Firstly, we reproduced ultraquantum and quasiclassical regimes at short
and long injection times
(in a related  calculation\cite{Yamamoto},
vortex injection was not strong enough to generate large length scales;
this is consistent with the facts that 
turbulence decayed as $L \sim t^{-1}$ and the relative forcing
\cite{forcing} was ten times less than in our first simulation).
Secondly, by computing the spectrum, we discovered \cite{Baggaley-ultra}
that the quasiclassical regime is the decay of a Kolmogorov spectrum, which
forms as energy is transferred from the small injection length scale to larger
length scales. However, the nonuniformity of the beam  
was (at least in principle) a possible origin of the observed inverse energy
transfer. 

In summary, the simulations which we present here, together with our 
previous result \cite{Baggaley-ultra},
show clearly the phenomenon of energy transfer from small to large scales 
and its relation with vortex reconnections. The negative energy flux
is observed not only in transients but also  
in statistically steady state regimes, and occurs over a wide range
of wavenumbers. 
The effect which we describe has implications
for other superfluid turbulence experiments, in particular
for the formation of developed turbulence past a grid, as in the towed-grid
experiments by Donnelly and collaborators \cite{Smith1993} which have
been much discussed in the literature, the oscillating
grid experiments performed
at the University of Lancaster~\cite{Bradley2005,Bradley2006},
and the most recent experiments of Walmsley et al. \cite{Walmsley2013}
which seem to confirm the inverse transfer of energy which we have
identified.

The analogies with related processes in classical fluid dynamics
are also intriguing, but need further detailed investigation. 
It is interesting to recall recent work by
Biferale et al. \cite{Biferale} who
numerically induced the classical three-dimensional inverse energy cascade 
by artificially restricting
the nonlinearity of the governing Navier-Stokes equation to
the interaction of Fourier modes of the same helical sign.
Their result shows that, in principle, all three-dimensional 
turbulent flows contain nonlinearities which may lead to an inverse 
cascade: to make the effect apparent one has to break the mirror 
symmetry of the interactions. Our findings are apparently 
consistent with Biferale's.
It must be stressed that it is not the anisotropy of the configuration 
which matters, but rather the anisotropy of the interaction.
In Biferale's problem, the anisotropy is enforced at every time
step by the numerical algorithm; in our problem, the
anisotropy is introduced by the initial condition
which favours one kind of
vortex reconnections over the other, as we have described.
In more isotropic conditions, the direct cascade generally 
may hide this effect (for example, a small inverse
energy transfer is apparent in the energy spectrum of a decaying
Taylor-Green flow~\cite{Araki-taylorgreen} although the authors 
do not comment on it).

In conclusion, the natural question is whether
the inverse energy transfer which we have described
amounts to a cascade, or creates an equilibrium
distribution at large scales as described for example
in the simpler case of wave turbulence \cite{Balkovsky}.
In a nonlinear system we expect that excitation in
a spectral interval means transfer of energy to both larger and smaller scales.
In our case, the interaction between vortex loops
is not symmetric and far from trivial. Numerical
simulations over a much wider range of wavenumbers and temporal scales
than we can perform now will help answering this question.

We acknowledge fruitful discussions with W.F. Vinen, L. Skrbek, A. Golov,
J. Laurie, P. Clark and S. Nazarenko,
and the financial support of the Leverhulme Trust, the EPSRC 
and the Carnegie trust.


\begin{thebibliography}{99}
\bibitem{angle}
The opening angle of the tree-algorithm is $0.3$.
\bibitem{spectrum}
Spectra are computed from a $512^2$ Cartesian mesh in the $xz$-plane.
\bibitem{forcing}
Let $D$ be the unit of length and $\kappa/\ell$ the unit of speed.
The unit of time is then $\tau=D\ell/\kappa$. 
In a characteristic time scale $\tau$, the injected vortex length
relative to the total vortex length (using length as a proxy for energy
for simplicity),
$\delta \Lambda/\Lambda=2 \pi \bar{R} f \tau/\Lambda$, is almost three
times larger in the fourth simulation than in the first.
\end{thebibliography}
\end{document}